\documentclass{article}

\usepackage{amsthm,amsmath,stmaryrd,bbm,hyperref,geometry,color}
\usepackage[utf8]{inputenc}
\usepackage[english]{babel}
\usepackage{graphicx}
\usepackage{amsfonts,amssymb}
\usepackage{verbatim}
\usepackage{enumitem}
\usepackage{algorithm}
\usepackage{authblk}
\usepackage{algpseudocode}
\usepackage{multirow}
\usepackage{xcolor}
\usepackage[autostyle]{csquotes}
\usepackage{caption}
\usepackage{subcaption}
\usepackage{siunitx}
\DeclareSIUnit\angstrom{\text {Å}}
\usepackage{booktabs}

\usepackage[style=numeric, sorting=none, isbn=false, url=false, maxnames=6, minnames=1, date=year, backend=biber]{biblatex}
\newcommand{\po}{\left(}
\newcommand{\pf}{\right)}

\newcommand{\R}{\mathbb R}

\newcommand{\E}{\mathbb E}

\newcommand{\dd}{\text{d}}
\newcommand{\na}{\nabla}

\newcommand{\veps}{\varepsilon}

\newcommand{\EM}{\ensuremath}

\newcommand{\cL}{\EM{\mathcal{L}}}

\hypersetup{
    unicode=false,          
    pdftoolbar=true,        
    pdfmenubar=true,      
    pdffitwindow=false,    
    pdfstartview={FitH},
    citecolor=blue,
    colorlinks=true,       
    linkcolor=blue,
    filecolor=blue,      
    urlcolor=blue          
}

\title{Faster Molecular Dynamics with Neural Network Potentials via Distilled Multiple Time-Stepping and Non-Conservative Forces}
\date{\today}

\author[1]{Nicolaï Gouraud}
\author[2]{Côme Cattin}
\author[2,*]{Thomas Plé}

\author[2]{Olivier Adjoua}
\author[1,2]{Louis Lagardère}
\author[1,2,*]{Jean-Philip Piquemal}

\affil[1]{Qubit Pharmaceuticals, Advanced Research Department, 75014 Paris, France}
\affil[2]{Sorbonne Université, Laboratoire de Chimie Théorique, UMR 7616 CNRS, 75005 Paris, France}
\affil[*]{Contact authors: thomas.ple@sorbonne-universite.fr, jean-philip.piquemal@sorbonne-universite.fr}
\begin{document}
\maketitle

\begin{abstract}
Following our previous work (J. Phys. Chem. Lett., 2026, 17, 5, 1288–1295), we propose the DMTS-NC approach, a distilled multi-time-step (DMTS) strategy using non-conservative (NC) forces to further accelerate atomistic molecular dynamics simulations using foundation neural network models such as FeNNix-Bio1. There, a dual-level reversible reference system propagator algorithm (RESPA) formalism couples a target accurate conservative potential to a simplified distilled representation optimized for the production of non-conservative forces. Despite being non-conservative, the distilled architecture is designed to enforce key physical priors, such as equivariance under rotation and cancellation of atomic force components. These choices facilitate the distillation process and therefore improve drastically the robustness of simulation, significantly limiting abnormal discrepancies between the two models, thus achieving excellent agreement with the forces data. Overall, the DMTS-NC scheme is found to be more stable and efficient than its conservative counterpart with additional speedups reaching 15-30\% over DMTS. Requiring no fine-tuning steps, it is easier to implement and can be pushed to the limit of the systems physical resonances to maintain accuracy while providing maximum efficiency.  We obtain additional speedup by combining hydrogen mass repartitioning (HMR), High Hydrogen Friction (HHF) to further extended the largest timestep up to 10fs of our schemes while conserving stability and accuracy. As for DMTS, DMTS-NC is applicable to any neural network potential and can be applied to approaches that are computationally heavier than FeNNix-Bio1. We show a proof of principle applying the approach to the distillation of MACE-OFF23 with consequent speedups ranging from $3.66$ to $5.64$ compared to single timestep.
\end{abstract}

\section{Introduction}

Molecular dynamics (MD)~\cite{Leimkuhler2015, Allen2017} is a popular numerical tool aimed at inferring properties of matter from computer simulations at the atomistic level. With the steady development of efficient algorithms and computational power, \textquote{in silico} methods enable the simulation of increasingly complex systems with high accuracy, driving a variety of scientific applications in biology, chemistry, drug design and material sciences~\cite{Ciccotti2022}. Studying the time evolution of a system, i.e. resolving the associated equations of motion, requires a physical model encoding the interactions between particles. Ideally, at this level of description, this model should be grounded on quantum mechanics. However, solving numerically the electronic structure problem, with so-called \textit{ab initio} methods~\cite{Jensen2017}, remains computationally expensive, which limits their application to relatively small systems and on short time scales. A more coarse grained approach, realizing a good compromise between computational efficiency and modeling at the atomic scale, consists in using an empirical potential~\cite{Ren2003,Reif2012,Maier2015,Robertson2015,Huang2017,Mackerell2005} with a given functional form, involving \textit{a priori} unknown parameters that must be optimized to replicate a set of targeted thermodynamic or structural properties.

In recent years, a third approach has emerged, namely Neural Network potentials (NNPs)~\cite{Behler2007,Bartok2010,Shakouri2017,Smith2017,Schutt2017,Chmiela2018,Grisafi2019,Zubatyuk2019,Devereux2020,Unke2021,Gasteiger2021,Grisafi2021,Musaelian2022,Ple2023,Unke2024,Kabylda2025}, which can be seen as the application of machine-learning architectures to the analytic potential approach, learning an energy model from large \textit{ab initio} computation databases. Nowadays, several libraries such as SchnetPack~\cite{Schutt2023}, DeepMD-kit~\cite{Zeng2023}, MLAtom~\cite{Dral2024} or FeNNol~\cite{Ple2024} enable the implementation of NNPs, and large foundation models such as MACE~\cite{Batatia2022,Kovacs2025} and FeNNix-Bio1~\cite{Ple2025} allow general-purpose simulations, covering complete areas of applications of molecular dynamics~\cite{Batatia2022,Kovacs2025,Ple2025,Ple2023,Benali2025}. Several important advantages over traditional approaches can be highlighted. First, the large number of parameters, whose optimization is fully automated, make NNPs highly flexible and largely transferable across systems. Second, they inherently handle chemical reactions, which most classical empirical force fields don't. Finally, again thanks to the large number of parameters, NNPs can achieve close-to quantum-mechanical accuracy, only for a fraction of the computation cost of $\textit{ab initio}$ methods. However, they remain significantly more expensive to evaluate than traditional empirical potentials, which motivates the development of algorithms increasing the speed of simulation based on NNPs without loosing their high accuracy.

In a molecular dynamics simulation, the number of evaluations of the gradient of the potential energy per unit of physical time is inversely proportional to the time discretization step size $\delta$. The maximum value of this parameter which preserves stability and accuracy of the simulation is limited by the highest-frequency motions of the system, such as chemical bond vibrations. 
A well-known approach to increase the step size used in classical integrators such as BAOAB~\cite{Leimkuhler2013a,Leimkuhler2013b} is called multi-time-stepping~\cite{Tuckerman1992,Zhou2001,Lagardere2019} (MTS). When applied to an empirical force field model, it consists in treating the fast-varying short-range forces within a small step size $\delta$, while only applying the long-range, more regular forces at larger time step $\Delta = n\delta$. This effectively increases the speed of simulation by reducing the rate at which expensive long-range forces are computed. Note that although MTS simulations are stable and accurate for $\Delta$ several times larger than $\delta$, its maximum value is also bounded, this time by resonance phenomena~\cite{Skeel1993,Skeel2003} resulting from a coupling of internal frequencies of the molecular system with nonphysical periodicity created by the various time steps, which can induce numerical instabilities for large values of $\Delta$. Despite these limitations, for which several alternatives exist~\cite{Morrone2011,Albaugh2019,Leimkuhler2013c,Margul2016,Gouraud2025}, MTS methods such as RESPA~\cite{Tuckerman1992} or BAOAB-RESPA1~\cite{Lagardere2019} are widely adopted. 

MTS schemes are also applicable to \textit{ab initio} MD~\cite{Liberatore2018}, and several recent works studied their use in machine learning approaches~\cite{Inizan2023,Fu2023,Osadchey2025,Mouvet2025,Cattin2026}. Note that the principle of MTS as described above is not readily usable with NNPs, since there is no natural decomposition of the forces into \textquote{cheap} and \textquote{expensive} parts. Recently, we proposed~\cite{Cattin2026} a strategy based on distillation~\cite{Hinton2015,Gou2021} of the FeNNix-Bio1(M) foundation model~\cite{Ple2025} which was performed via the FeNNol~\cite{Ple2024} library. In this strategy, a small, less precise but fast-to-evaluate model is trained on data labeled with the FeNNix-Bio1(M) model instead of Density Functional Theory. Then, following the MTS procedure, this small model is applied in several iterations of an inner loop, before being corrected by an application of the difference between the large and small force models in an external loop. Therefore, by reducing the number of computations of the large expensive model, simulation speed is increased, enabling longer stable simulations.

As proposed in several previous works~\cite{Gasteiger2021,Hu2021,Liao2023,Neumann2024,Rhodes2025,Eissler2025}, another path to accelerating simulation speed with ML models is to use non-conservative forces, i.e. forces that do not necessarily derive from a potential. Indeed, directly predicting interatomic forces enables to bypass the costly backpropagation step required to compute conservative forces, thus improving computational efficiency at fixed model architecture. However, Bigi et al.~\cite{Bigi2025a} identified severe simulation artifacts induced by this approach and alternatively proposed to use non-conservative forces in an MTS scheme. In their framework, a single model is trained to output both conservative and non-conservative forces: the latter are used at every step of the simulation, while conservative forces are only evaluated every few steps. In this work, we follow a similar direction and extend our Distilled Multi-Time-Step (DMTS) procedure to the use of non-conservative forces, which we denote DMTS-NC. Compared to the unified model proposed in~\cite{Bigi2025a}, the distillation step significantly reduces the computational cost of the model used in the MTS internal steps, while the absence of differentiation procedures make both training and evaluation faster than the distilled conservative models used previously in~\cite{Cattin2026}. In addition, by imposing several physical priors such as equivariance under rotation and cancellation of atomic force components (which are properties that are satisfied by conservative forces), the distilled model is able to closely fit the reference one. We show that the improved agreement between the models, with training reaching a mean absolute error near 1 kcal/mol/\AA\ on a diverse dataset labeled with FeNNix-Bio1(M), allows to use larger external time steps, further enhancing the speed and robustness of DMTS schemes. Importantly, like the original DMTS framework, DMTS-NC is a model-agnostic acceleration strategy, and can be applied to any neural network potential, independently of its specific architecture.

The rest of the article is organized as follows. In Section~\ref{sec:methods}, we describe the architecture and training of the non-conservative model, the associated DMTS-NC integrator, as well as several techniques used to further stabilize the dynamics, namely Hydrogen Mass Repartitioning (HMR), High Hydrogen Friction (HHF) and a rewind procedure preventing occasional disagreements between the models. Section~\ref{sec:numerics} is devoted to numerical experiments and validation of the improved stability and speed of DMTC-NC on various molecular systems. In addition to simulations based on FeNNix-Bio1, we highlight the generality of DMTS-NC by applying the method to the MACE-OFF23 foundation model.

\section{Methods}\label{sec:methods}

\paragraph{Model architecture.}

The architecture of the non-conservative model is a modification of FeNNix-Bio1~\cite{Ple2025} in which the neural network learns to directly predict forces instead of energies. By removing the requirement of having conservative forces, i.e deriving from a potential, certain properties such as Newton's third law are not guaranteed anymore. However, our architecture is designed to still enforce some physical priors, such as equivariance under rotation and cancellation of atomic force components. 

Let us first quickly recall the general architecture of FeNNix-Bio1. In a first step, each atom is embedded in a $N_f$-dimensional vector space, describing electronic structure and charge information. This initial geometry-independent embedding is then updated to include local geometric information via two message-passing layers of an equivariant transformer.
The first layer considers a high-resolution description of very short range (within a $R_c^{(sr)}=\SI{3.5}{\angstrom}$ cutoff) geometry, while the second layer incorporates both short-range and medium-range messages up to a cutoff $R_c^{(lr)}=\SI{7.5}{\angstrom}$. Then, atomic energies are obtained from mixture-of-experts multi-layer perceptrons that take the final embeddings as an input, with routing depending on the chemical group of each atom. Finally an explicit screened nuclear repulsion term is added. We refer to~\cite{Ple2025} for more details.

We now turn to the non-conservative model. Its embedding architecture is similar to the previous one, except that only short-range messages are considered in both layers. This allows the model to mainly focus on local structure, as it is specifically designed to handle short-range forces and the associated high-frequency motions within the inner loops of an MTS integrator. At the end of the embedding layers, each atom $i$ is characterized by a scalar embedding $x_i\in\R^{N_f}$ and a tensorial embedding $\hat{V}_i\in\R^{{n_l}\times(\lambda_{\max}+1)^2}$. The latter is a collection, along $n_l$ channels, of geometric tensors of orders up to $\lambda_{max}$. We then use $n_s$ scalar attention heads and $n_l$ tensor heads, one for each first-order geometric tensor channel. For every atom $i$ and each attention head $h\in\{1,\dots,n_s+n_l\}$, vectors $q_{ih}$ and $k_{ih}$ are computed via a linear projection of the scalar embedding $x_i$ on a subspace of dimension $n_c$:
\begin{align*}
    q_{ih} &= W_{qh} x_i \\
    k_{ih} &= W_{kh} x_i\,,
\end{align*}
where the weight matrices $W_{qh}$ and $W_{kh}$ are optimized during the training. Let $\mathcal{N}(i)$ denote the neighbor list of atom $i$, i.e. the set of atoms $j\neq i$ within the short-range cutoff distance $R_c^{(sr)}$. For each head $h\in\{1,\dots,n_s+n_l\}$ and each neighbor $j\in\mathcal{N}(i)$, the scaled dot product coefficient $c_{ijh}$ is defined as
\[c_{ijh} = \frac{1}{\sqrt{n_c}}\sum_{k=1}^{n_c}q_{ihk}k_{jhk}\,.\]
Let $\vec{R}_{ij}$ be the vector pointing from $i$ to $j$ and $(B_h(r_{ij}))_{h\in\{1\dots n_s\}}$ a radial basis vector
corresponding to the decomposition of $r_{ij}=|\vec{R}_{ij}|$ in a basis of $n_s$ Bessel functions, as described in~\cite{Gasteiger2020}. The three-dimensional vector $F_{ij}$ is then constructed as
\[F_{ij} = \po \sum_{h=1}^{n_s} c_{ijh} B_h(r_{ij})\pf \po\frac{\vec{R}_{ij}}{r_{ij}}+\sum_{h=1}^{n_l} c_{ij(n_s+h)}\vec{V}_{ih}\pf f_c(r_{ij})\,,\]
where $\vec{V}_{ih}$ is the vector part of the tensor embedding $\hat{V}_{ih}$ (i.e. the irreducible representations of order $\lambda=1$), $f_c(r_{ij})$ is a polynomial cutoff function going smoothly to zero at the short-range cutoff distance $R_c^{(sr)}$, as defined in~\cite{Gasteiger2020}. The final force vector $F_i$ acting on atom $i$ is given by the antisymmetric sum of the $F_{ij}$ to which is added the short-range screened nuclear repulsion term $F_{\text{rep}} = -\na E_{\text{rep}}$ that is used in FeNNix-Bio1:
\[ F_i = \sum_{j\in\mathcal{N}(i)}(F_{ij}-F_{ji}) + F_{rep,i}\in\R^3\,.\]
The expression of $E_{\text{rep}}$ follows the analytic NHL parametrization from~\cite{Nordlund2025}:
\[E_{rep}(r_{ij}) = \frac{Z_iZ_j}{4\pi\veps_0r_{ij}}\sum_{n=1}^3 a_{nij}e^{-b_{nij}r_{ij}}\,,\]
where $Z_i$ is the atomic number of atom $i$, and $a_{nij},b_{nij}$ are parameters, obtained via fitting to \textit{ab initio} reference data in~\cite{Nordlund2025}, that depend on the unordered pair of species $Z_i,Z_j$.
 
Note that $F_i$ explicitly depends on the tensorial embedding $\hat{V}_i$, while in the original architecture of FeNNix-Bio1 it is only used as an internal descriptor in the embedding phase, and not given as an input to the perceptrons that compute the final energies. In particular, here the functional form of $F_i$ ensures that the sum of the forces over all atoms is zero, and more generally the sum of the $F_i$ over any connected component of the interaction graph of the atoms is also zero. In addition, all the $F_{ij}$, and therefore the $F_i$ too, are equivariant under rotation.

We conclude this paragraph with the parameter values. In FeNNix-Bio1(M), $N_f=176$, $n_c=16$, $\lambda_{max}=3$, $n_l=4$ and $n_s = 10$. In the small NC model, we have $N_f = 64$, $n_c=8$, $\lambda_{max} = 1$, $n_l=2$ and $n_s=8$. For the sake of completeness, let us mention that in the initial geometry-independent embedding, the number of charge equilibration channels is $N_Q = 32$ in FeNNix-Bio1(M), and $N_Q= 8$ in the NC model, the species embedding neural network involves one layer of 256 neurons in FeNNix-Bio1(M) and only 64 in the NC model. In the message-passing phase, the attention mechanism involves $h_s=16$ scalar attention heads in FeNNix-Bio1(M) and $h_s=4$ in the NC model, and a two hidden-layers neural network, with respectively 352 and 176 neurons in FeNNix-Bio1(M) and a one-hidden-layer neural network with 64 neurons in the NC model. Finally, the polynomial $f_c(r_{ij})$ is of order 8 in FeNNix-Bio(M), and 5 in the NC model. This results in a total number of 286 736 parameters in the NC model, compared to 9 526 855 for FeNNix-Bio1(M).

\paragraph{Model training.}

Following the principle of knowledge distillation~\cite{Hinton2015,Gou2021}, as applied in~\cite{Cattin2026}, the non-conservative force model is trained as a distilled version of the larger FeNNix-Bio1(M)~\cite{Ple2025} NNP. A subset of conformations of the SPICE2 dataset~\cite{Eastman2023,Eastman2024} is evaluated with FeNNix-Bio1(M), which then constitutes the training dataset for the non-conservative force model. This dataset contain a variety of small organic molecules and biologically relevant complexes, thus covering a wide chemical space.

The training was done for $2200$ epochs with $1000$ batches of $128$ conformations per epoch. We used the Muon optimizer~\cite{Jordan2024}, with an initial learning rate of $1.0\times10^{-5}$, linearly increasing up to $5.0\times10^{-4}$ after a warm-up phase, then decreasing down to a final rate of $1.0\times 10^{-6}$ according to a cosine one-cycle learning-rate schedule. The total training, performed on a Nvidia A100 40GB GPU, took 28 hours and 47 minutes. The resulting model, which directly aims at fitting the force vectors produced by FeNNix-Bio1(M) instead of the energies, ends up achieving excellent agreement with the data, reaching a final mean absolute error of MAE=$1.46$kcal/mol and a root mean square error of RMSE=$2.33$kcal/mol, a significantly lower value than for the small conservative model of~\cite{Cattin2026}, for which MAE=$3.44$kcal/mol and RMSE=$5.53$kcal/mol.

\paragraph{Multi-time-step integrator.}

Denote $U:\R^d\rightarrow \R$ a potential energy function, and $\cL$ the infinitesimal generator of associated Langevin dynamics. Most single-time-step (STS) numerical schemes rely on so-called splitting of the dynamics. For instance, the BAOAB method~\cite{Leimkuhler2013a,Leimkuhler2013b} consists in splitting the Langevin process into three parts A, B and O, corresponding respectively to free transport, acceleration and fluctuation/dissipation, which comes down to splitting the generator as $\mathcal{L} = \mathcal{L}_A+\mathcal{L}_B+\mathcal{L}_O$ and simulating successively the parts B-A-O-A-B, thanks to the Trotter/Strang formula
\[e^{t\mathcal L} = e^{\frac t2\mathcal L_B}  e^{\frac t2\mathcal L_A} e^{t\mathcal L_O} e^{\frac t2\mathcal L_A} e^{\frac t2\mathcal L_B} + \underset{t\rightarrow 0}{\mathcal O}(t^3)\,.\]
Multi-Time-Step (MTS) methods such as BAOAB-RESPA push the splitting of the dynamics one step further, by decomposing the force vector field $-\na U = F_{S}+F_{L}$ and considering separately the associated accelerations $\mathcal L_{B_S}$ and $\mathcal L_{B_L}$. In a classical force field, $F_S$ typically gathers the fast-varying, easy to compute short-range forces and $F_L$ includes more regular many-body or long-range forces that are more expensive to evaluate.

Given a time step $\delta>0$, denote $Q_\delta$ one transition of the BAOAB chain associated with the force $F_S$, namely

\[
Q_\delta  =  e^{\frac \delta 2\mathcal L_{B_S}} e^{\frac \delta  2\mathcal L_A} e^{\delta  \mathcal L_O} e^{\frac \delta  2\mathcal L_A} e^{\frac \delta 2\mathcal L_{B_S}}\,.\]

Then, by denoting $\Delta = n\delta$ some multiple of $\delta$, the BAOAB-RESPA MTS scheme relies on the following approximation of the continuous dynamics:

\[ e^{\Delta\cL} \approx e^{\frac\Delta2\cL_{B_L}} (Q_\delta)^n e^{\frac\Delta2\cL_{B_L}} \,.\]
In other words, letting $x$ and $v$ denote the position and velocity vectors and $M$ the mass matrix, one transition of the integrator consists of the following steps:

\begin{enumerate}
    \item $v\leftarrow v-\frac{\Delta}{2}M^{-1}F_L(x)$\,.
    \item Perform $n$ times the BAOAB scheme associated with $F_S$ with time step $\delta$\,.
    \item $v\leftarrow v-\frac{\Delta}{2}M^{-1}F_{L}(x)$\,.
\end{enumerate}

Note that in this method, $F_S$ and $F_L$ are not required to be conservative forces: they only need to satisfy $F_S+F_L = -\na U$. In DMTS-NC, $F_S$ is the smaller, faster to evaluate machine-learned model approximating $\na U$ that we described above and $F_L=-\na U-F_S$. To sum up, the full procedure is given in Algorithm 1.

\begin{algorithm}[htbp]
\caption{DMTS-NC Integration Step \\
\textbf{Input: } $\text{FENNIX}_\text{large}(x)$ the reference force field evaluated at configuration $x$, $\text{FENNIX}_\text{NC}(x)$ the cheaper non-conservative force model, $M$ the mass matrix, $\Delta$ the external time-step, $n$ the number of iterations of the internal loop.}
\label{algo:baoab-respa}
\begin{algorithmic}[1]
\If{$\text{first\_step}$}
\State $F_\text{NC} \gets \text{FENNIX}_\text{NC}(x)$ 
\State $F \gets \text{FENNIX}_\text{large}(x)$
\EndIf
\State $v \gets v + \dfrac{\Delta }{2} M^{-1}(F - F_\text{NC})$
\For{$i = 1$ to $n$}
    \State $v \gets v + \dfrac{\Delta}{2n} M^{-1}F_\text{NC}$
    \State $x \gets x + \dfrac{\Delta}{2n}v$
    \State $v \gets \text{Thermostat}\po v,\dfrac{\Delta}{n}\pf$
    \State $x \gets x + \dfrac{\Delta}{2n}v$
    \State $F_\text{NC} \gets \text{FENNIX}_\text{NC}(x)$
    \State $v \gets v + \dfrac{\Delta}{2n}M^{-1}F_\text{NC}$
\EndFor
\State $F \gets \text{FENNIX}_\text{large}(x)$
\State $v \gets v + \dfrac{\Delta}{2}M^{-1}(F - F_\text{NC})$
\end{algorithmic}
\end{algorithm}

\paragraph{Hydrogen Mass Repartitioning.}

The resonance phenomenon~\cite{Skeel1993,Skeel2003} that limits the maximum usable value of $\Delta$ can be mitigated in various ways. A standard procedure, applied for instance in~\cite{Lagardere2019}, and which doesn't significantly impair sampling rates is called Hydrogen Mass Repartitioning (HMR)~\cite{Feenstra1999}. It consists in increasing the mass of hydrogen atoms (in our simulations, by $3.0$ Daltons) and redistributing the deficit over heavier atoms in order to preserve the total mass of each molecule. This has the effect of red-shifting the frequency of intramolecular bond and angle motions, allowing in turn to increase the external time step $\Delta$, without biasing configurational sampling (that only depends on potential energy, which is unchanged), nor the temperature (whose distribution doesn't depend on the mass). In addition, as shown below in Section~\ref{sec:numerics} and already noted in~\cite{Lagardere2019,Cattin2026}, by preserving the mass of each molecule, dynamic properties such as diffusion coefficients are not significantly impacted either, contrary to other methods~\cite{Albaugh2019,Leimkuhler2013c,Margul2016}.

\paragraph{High Hydrogen Friction.}

To sample the canonical measure, a standard choice is to simulate (underdamped) Langevin dynamics, parametrized by a friction coefficient $\gamma$ measuring the rate of velocity dissipation. This process can be generalized by replacing the scalar $\gamma$ by any (possibly position-dependent) symmetric positive-definite matrix $\Gamma$. While determining an optimal $\Gamma$ remains a non-trivial question (see~\cite{Chak2023} for results in this direction), a simple possibility is to take a constant diagonal matrix, assigning a specific friction coefficient to each atomic species. 

Taking a higher friction on hydrogen (a method to which we refer as HHF in the following) has the effect of damping bond oscillations, without impacting the diffusion coefficient (see Section~\ref{sec:numerics} below) as much as a fully overdamped regime on all atoms, while also being much simpler to implement and to parametrize than colored-noise thermostats based on the generalized Langevin equation~\cite{Morrone2011}.

In addition, and in combination with HHF, the intrinsically diffusive behavior caused by momentum flips in overdamped regimes can be addressed by replacing the standard Langevin thermostat (or rather its species-dependent-friction variant) by the Fast Forward Langevin (FFL) thermostat~\cite{Hijazi2018}. It consists in applying the following velocity update in the thermostat part of the integrator: 
\[v\gets \left|e^{-\gamma\delta}v+\sqrt{\beta^{-1}(1-e^{-2\delta\gamma}})M^{-1/2}G\right|\frac{v}{|v|}\,,\]
where $G$ is a standard multidimensional Gaussian vector, $M$ the mass matrix of the system, $\beta$ the inverse temperature, $\gamma$ the (species-dependent) friction coefficient and $|\cdot|$ the Euclidean norm.

\paragraph{Preventing rare model disagreements (Rewind procedure).}

In addition to intrinsic resonance problems associated with multi-time-stepping, DMTS faces a specific pitfall: the possibility that on rare occasions, $F_S$ and $-\na U$ disagree significantly, which can make the dynamics suddenly diverge because of the abnormally strong applied force $F_L = -\na U - F_S$.  Unlike resonances, which are detectable early in a simulation, these failures caused by \textquote{hallucinations} of the small force field are difficult to anticipate, since the dynamics can remain stable for a long period before encountering a disagreement.

One way of addressing this issue, proposed in~\cite{Cattin2026}, relies on an active learning procedure. It consists in fine-tuning the small model based on data frames, collected automatically during an initial simulation, for which the difference between the models exceeds some unrealistic threshold. In addition to being system-specific, collecting the pathological data can take a long time if disagreements are too rare, in which case extremely long simulations are also required after the fine-tuning to validate that the new model is fully stable. As we show in Section~\ref{sec:numerics}, the excellent overall agreement between models, allowed by the non-conservative architecture focusing on forces instead of energies, results in much fewer \textquote{holes} than in the original DMTS of~\cite{Cattin2026}. Therefore, even at larger values of $\Delta$, the dynamics remains robust without relying on system-specific fine-tuning.

Instead, in DMTS-NC, a generic on-the-fly procedure is implemented: at each time step, if the difference between the two force models (or some other indirect indicator) exceeds a threshold, the dynamics rewinds until the last saved frame and MTS is replaced by STS for a given number of steps (typically a few picoseconds of physical time), before going back to multi-time-stepping. Given the rarity of those rewinds, occurring on average every $10$ns at the maximum stable $\Delta$, the overall simulation speed is almost not impacted by the procedure.

\paragraph{Computation of the virial term in NPT.} 

Several barostats are currently implemented in FeNNol for simulations in the isothermal-isobaric (NPT) ensemble: Monte-Carlo~\cite{Chow1995,Aqvist2004}, Berendsen~\cite{Berendsen1984}, Langevin piston~\cite{Feller1995} and Bussi's stochastic velocity rescaling~\cite{Bussi2009}.

A common issue of multi-time-step methods in NPT is that the small discretization bias on the potential energy, while insignificant in NVT for most observables of interest, has a non-negligible impact on the estimation of the virial term $q\cdot\na U(q)$, which, in turn, induces a problematic bias on the volume distribution. In the Tinker molecular dynamics software~\cite{tinker8}, this virial bias is corrected by adding the average of the intermediate virials at the end of each internal loop to the final estimator, computed from the configuration and volume at the end of the external step.

Here we apply a similar method, however since non-conservative forces are applied in the internal loops, the intermediate virials are not well defined. Instead, the correction term added to the final estimator of the virial is the average, over all internal loops, of the proxy term $-\sum_{i=1}^N F_i\cdot q_i$, where $F_i$ is the force acting on atom $i$. To resolve the coordinate ambiguity of $q_i$ regarding the choice of the periodic image under periodic boundary conditions, this term can be rewritten as a function of the direction vectors $r_{ij}=q_j-q_i$ with minimum image convention, and the pairwise force terms $F_{ij}$ and $F_{rep,i,j}$ defined in the model architecture section. By using the fact that the repulsion terms are anti-symmetrical and that neighbor lists are symmetrical, a simple computation shows that
\[-\sum_{i=1}^N F_i \cdot  q_i = \sum_{i=1}^N\sum_{j\in \mathcal{N}(i)} (F_{ij}+\frac{1}{2}F_{rep,i,j})\cdot r_{ij}\,.\]
In Section 1 of Supporting Information, we show that applying this correction term removes the bias on pressure and density distributions.

\section{Numerical results}\label{sec:numerics}

The goal of this section is to assess performance, stability and sampling accuracy of the DMTS-NC method, on bulk water boxes of various sizes, two solvated proteins and a collection of solvated small molecules. 
Simulations were performed in periodic boundary conditions, applying the Langevin thermostat at $T = 300$K temperature and $\gamma=\SI{1}{ps^{-1}}$ friction. In the HHF variant, we tested both Langevin and Fast Forward Langevin thermostats with a higher friction of $\gamma_H=10$ps$^{-1}$ on hydrogen atoms and $\gamma=\SI{1}{ps^{-1}}$ on other species. Single time step simulations used the BAOAB integrator. In all DMTS and DMTS-NC simulations, HMR was systematically activated with an additional $3.0$ Daltons on hydrogen atoms. Isobaric-isothermal simulations applied Bussi's barostat~\cite{Bussi2009}, targeting a $1.0$atm pressure, with a $\tau_P = 2.0$ps volume relaxation time. Training and MD used the FeNNol~\cite{Ple2024} Python library coupled to the Tinker-HP package \cite{lagardere2018tinker,adjoua2021tinker} within its Deep-HP module \cite{Inizan2023} on a single NVIDIA A100 40GB GPU card.

\subsection{Speed and sampling performances}\label{sec:perfs}

\begin{figure}[htbp]
    \centering
    \subfloat[Oxygen-oxygen radial distribution (1600 water molecules)]
    {\includegraphics[width=0.5\linewidth]{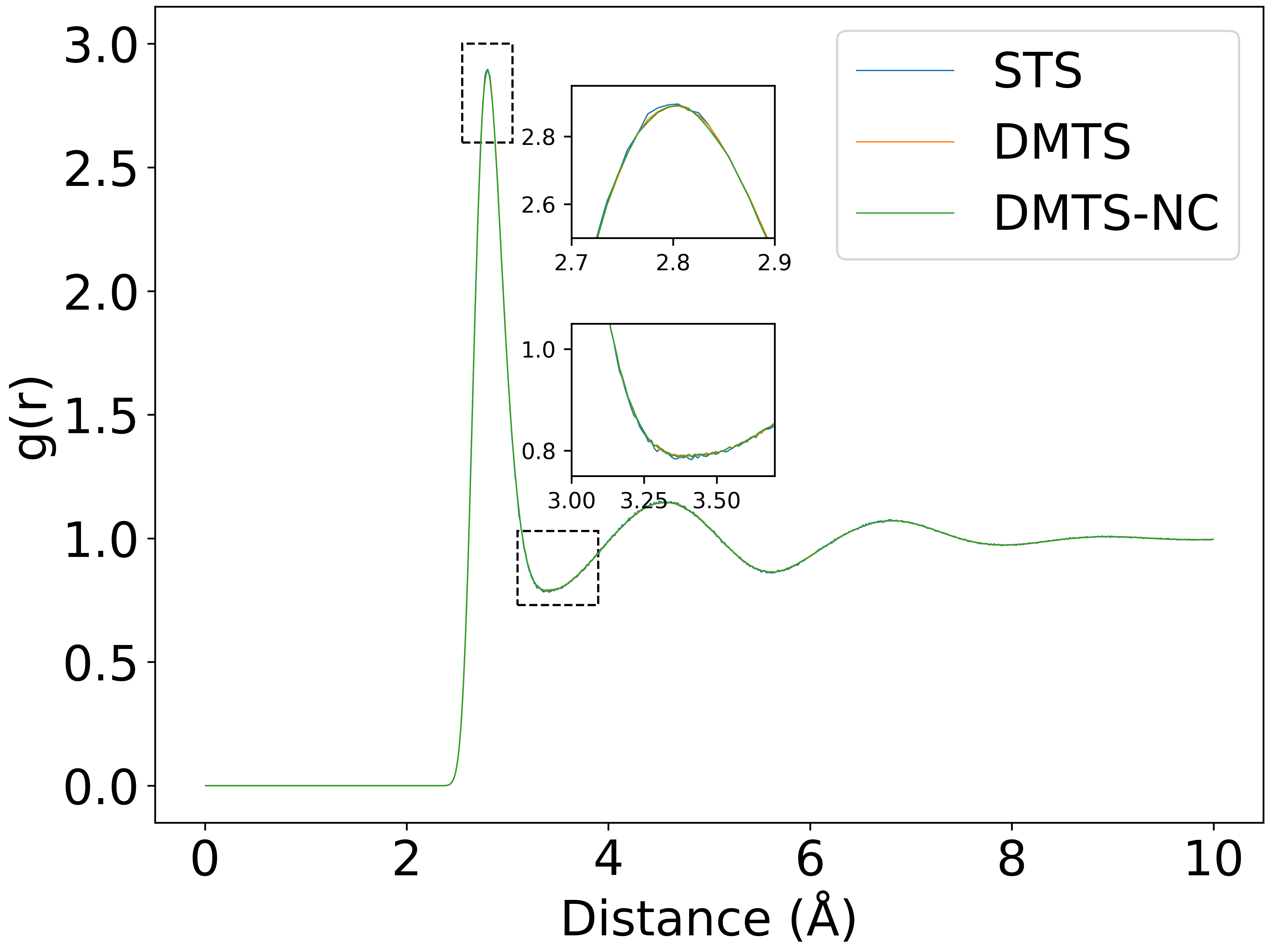}}
    \subfloat[Temperature distribution (216 water molecules)]{\includegraphics[width=0.5\linewidth]{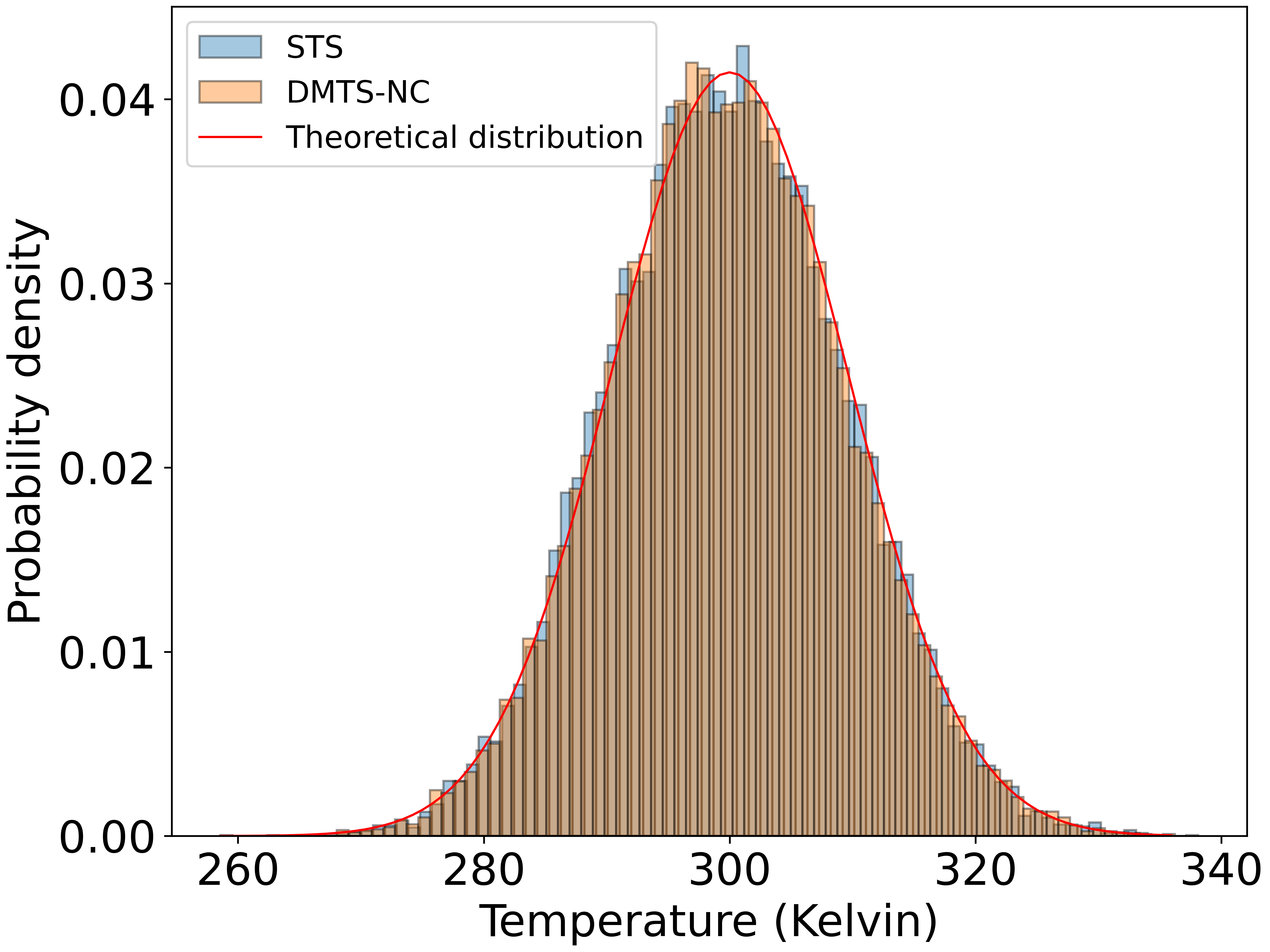}} \\
    \subfloat[Potential energy distributions: box plots]{\includegraphics[width=0.5\linewidth]{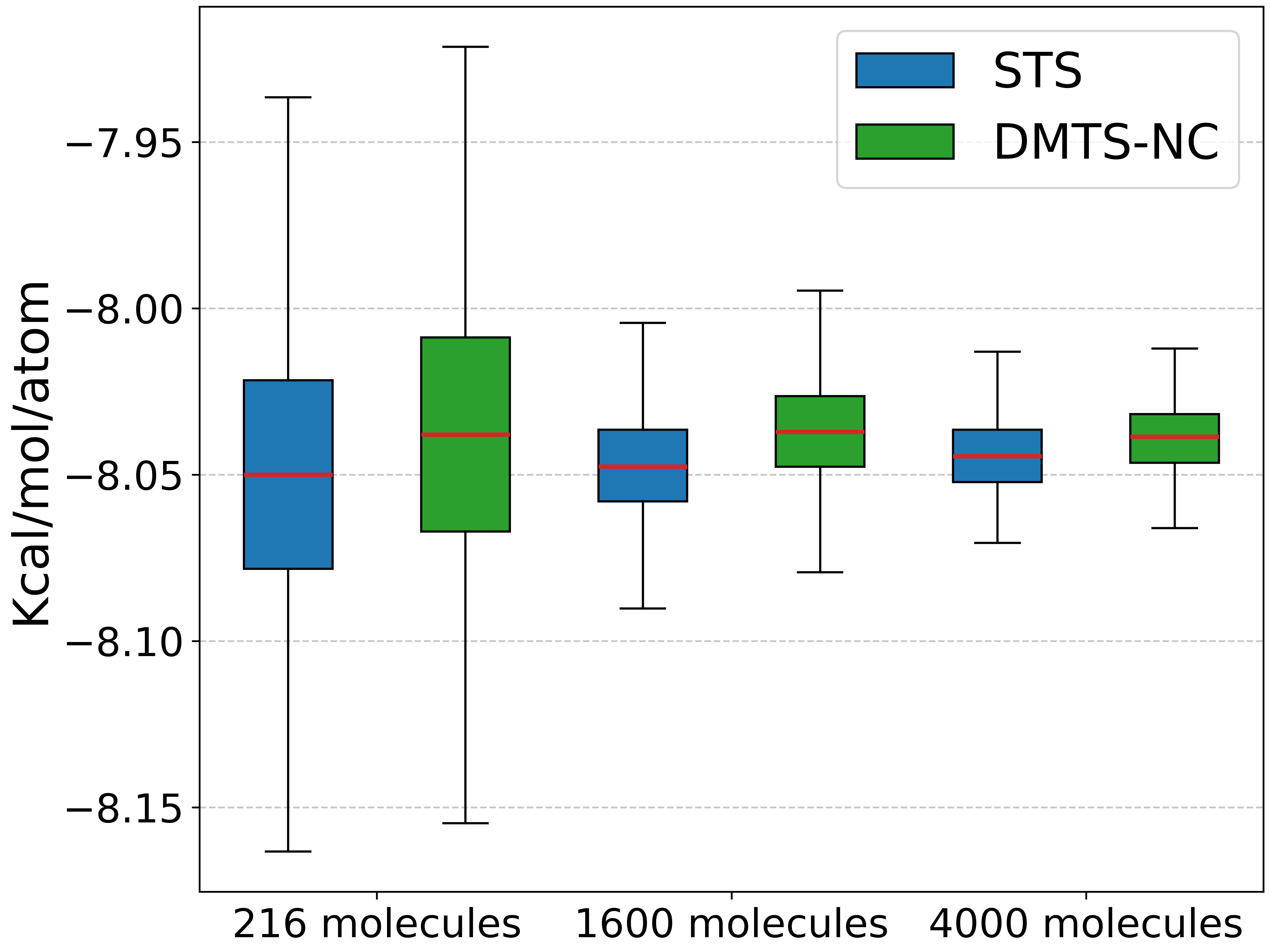}}
    \subfloat[Potential energy distributions: Q-Q plots]{\includegraphics[width=0.5\linewidth]{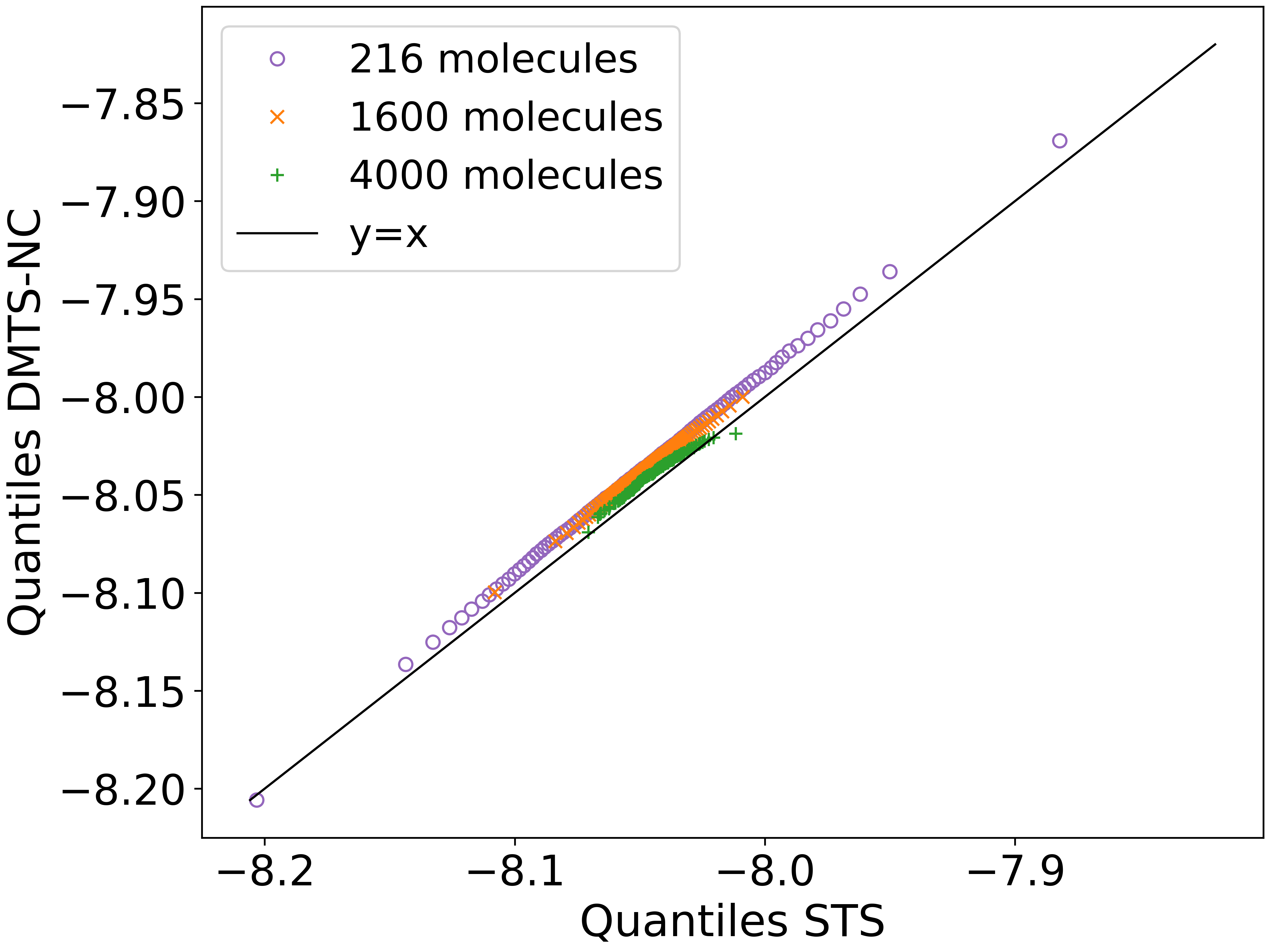}}
    \caption{Sampling properties of bulk water simulations. All data was collected from 2 nanosecond simulations, with one frame saved per picosecond. Figure (a) shows that radial distributions are not affected by the use of DMTS-NC compared to DMTS or STS, Figure (b) shows that temperatures fit the equilibrium theoretical distribution, and samples of potential energies, represented on a boxplot (Figure (c)) and a quantile-quantile plot (Figure (d)), highlight the low-bias identical-variance deviation of the distributions from STS to DMTS-NC, as expected for a multi-time-step method in a stable regime.}
    \label{fig:watersampling}
\end{figure}

\paragraph{Bulk water.} We start by comparing, for various sizes of bulk water boxes and various time steps, the STS BAOAB integrator, the BAOAB-RESPA DMTS integrator with the small conservative model used in~\cite{Cattin2026} and DMTS-NC, the version using the non-conservative force model. To assess sampling properties, Figure~\ref{fig:watersampling} shows oxygen-oxygen radial distributions (computed with FeNNol), as well as distributions of temperatures and potential energies across simulations. In particular, empirical distributions of temperature (known to be sensitive to numerical instabilities due to multi-time-stepping) are plotted along the equilibrium theoretical distribution, given by $\frac{T}{d}\chi^2(d)$, where $T$ is the equilibrium average temperature (here, \SI{300}{K}), $d$ is the total configurational dimension of the system (i.e. three times the number of atoms) and $\chi^2(d)$ is the Chi-squared distribution with $d$ degrees of freedom. In addition, Table~\ref{table:temp-epot} sums up the average potential energies and temperatures.

Table~\ref{table:waterspeed} shows the speed of the algorithms expressed in nanoseconds of simulation per day, highlighting the significant acceleration of DMTS-NC with respect to conservative model DMTS and to STS. Note that in addition to being faster, thanks to a better fit to the training data, DMTS-NC exhibits fewer numerical instabilities, and the maximum usable time step is limited by classical MTS resonances instead of the problem of abnormal model disagreements as described in Section~\ref{sec:methods}. Simulations show that DMTS-NC provides an acceleration ranging from $2.94$ to $4.31$ with respect to STS, and from $30.8\%$ to $55.7\%$ with respect to conservative DMTS.

\begin{table}[htbp]
\centering
\small 
\begin{tabular}{@{} l S[table-format=3.2] S[table-format=2.2] S[table-format=2.2] l @{}}
\toprule
& \multicolumn{3}{c}{\textbf{Number of atoms}} & \textbf{Numerical} \\
\cmidrule(lr){2-4}
\textbf{Integrator} & {648} & {4800} & {12000} & \textbf{instabilities} \\ 
\midrule
STS              & 36.89  & 9.08   & 3.63   & No \\
DMTS 1-5fs       & 69.54  & 30.17  & 11.95  & No \\
DMTS 1-6fs       & 80.62  & 34.34  & {NaN} & Data holes \\
DMTS 1-7fs       & {NaN}  & {NaN}  & {NaN}  & --- \\
\midrule
DMTS-NC 1-5fs    & 91.13  & 32.69  & 12.39  & No \\
DMTS-NC 1-6fs    & 103.67 & 36.15  & 13.67  & No \\
DMTS-NC 1.625-6.5fs & 108.29 & 41.10 & 15.64 & No \\
DMTS-NC 1-7fs    & 114.11 & 39.72  & 13.56  & High temp.\ (resonances) \\
\midrule
Max. acc. DMTS-NC/DMTS & {55.7\%} & {36.2\%} & {30.8\%} & \\
Max. acc. DMTS-NC/STS  & {2.94}  & {4.53}  & {4.31}  & \\
\bottomrule
\end{tabular}
\caption{Simulation performances for bulk water systems, expressed in nanoseconds per day. In DMTS-NC, the stability limit is reached around an external time step of $\Delta=6.5$fs, after what resonances start to appear. In contrast, the main limitation the with the (generic) conservative model is the presence of \textquote{data holes} already at $\Delta=6$fs.}
\label{table:waterspeed}
\end{table}

\paragraph{Solvated proteins.} 

We next turn to two protein systems: solvated phenol-lysosyme protein-ligand complex (16221 atoms) and solvated dihydrofolate reductase protein (DHFR, 23558 atoms). Simulation speeds are summarized in Table~\ref{table:proteinspeed}, showing that DMTS-NC provides an acceleration factor with respect to STS of $3.40$ for phenol-lysosyme and $3.09$ for DHFR, an acceleration with respect to the generic conservative DMTS of $31.6\%$ for phenol-lysosyme and $28.3\%$ for DHFR, and finally, an acceleration with respect to the fine-tuned (after a system-specific active learning procedure) conservative DMTS of $16.4\%$ for phenol-lysosyme and $15.1\%$ for DHFR. Note that in all simulations, DMTS-NC uses a generic, non-fine-tuned model.

\begin{figure}[htbp]
    \centering
    \subfloat[Protein backbone RMSD]
    {\includegraphics[width=0.5\linewidth]{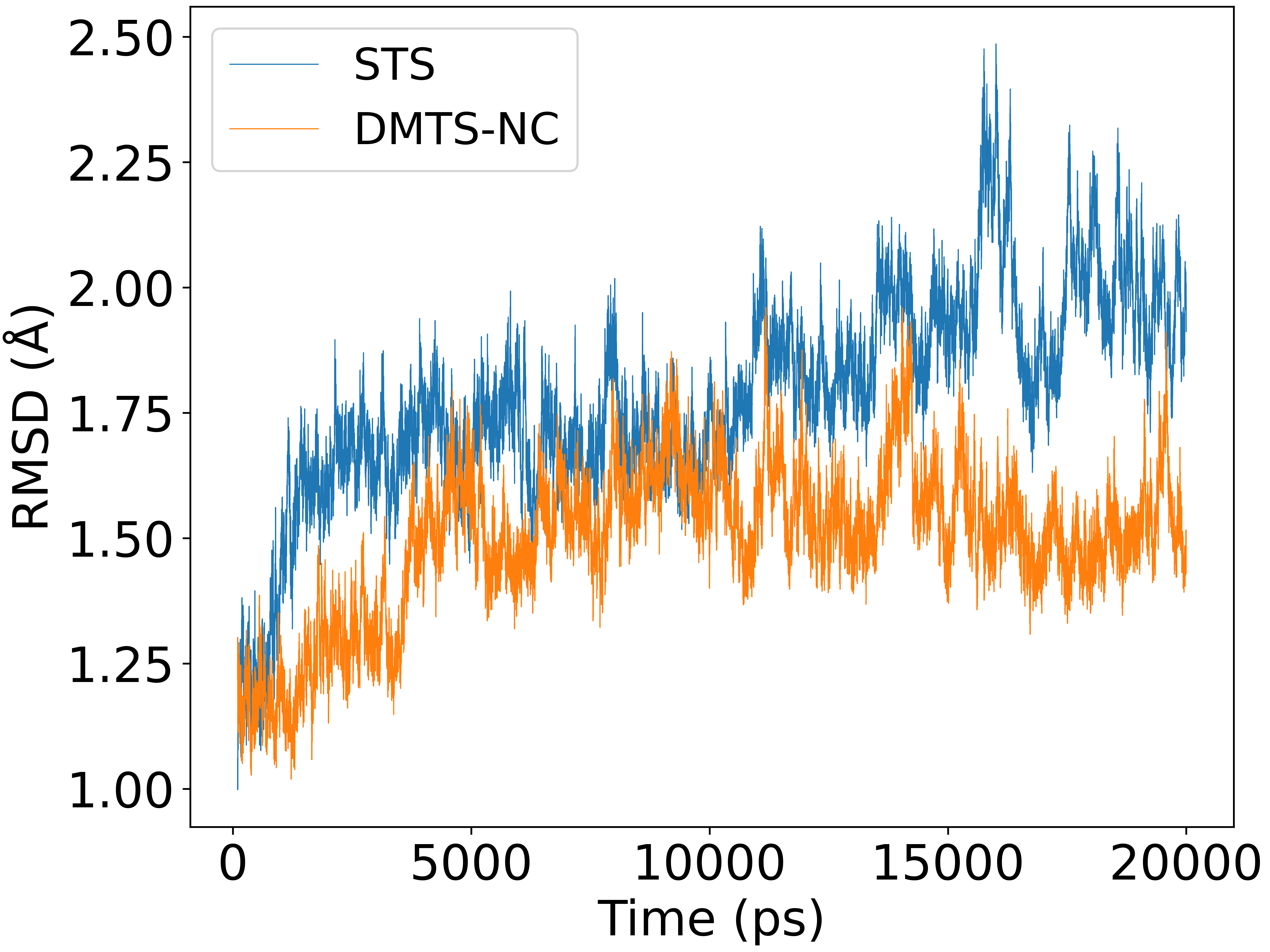}}
    \subfloat[Ligand DBC]{\includegraphics[width=0.5\linewidth]{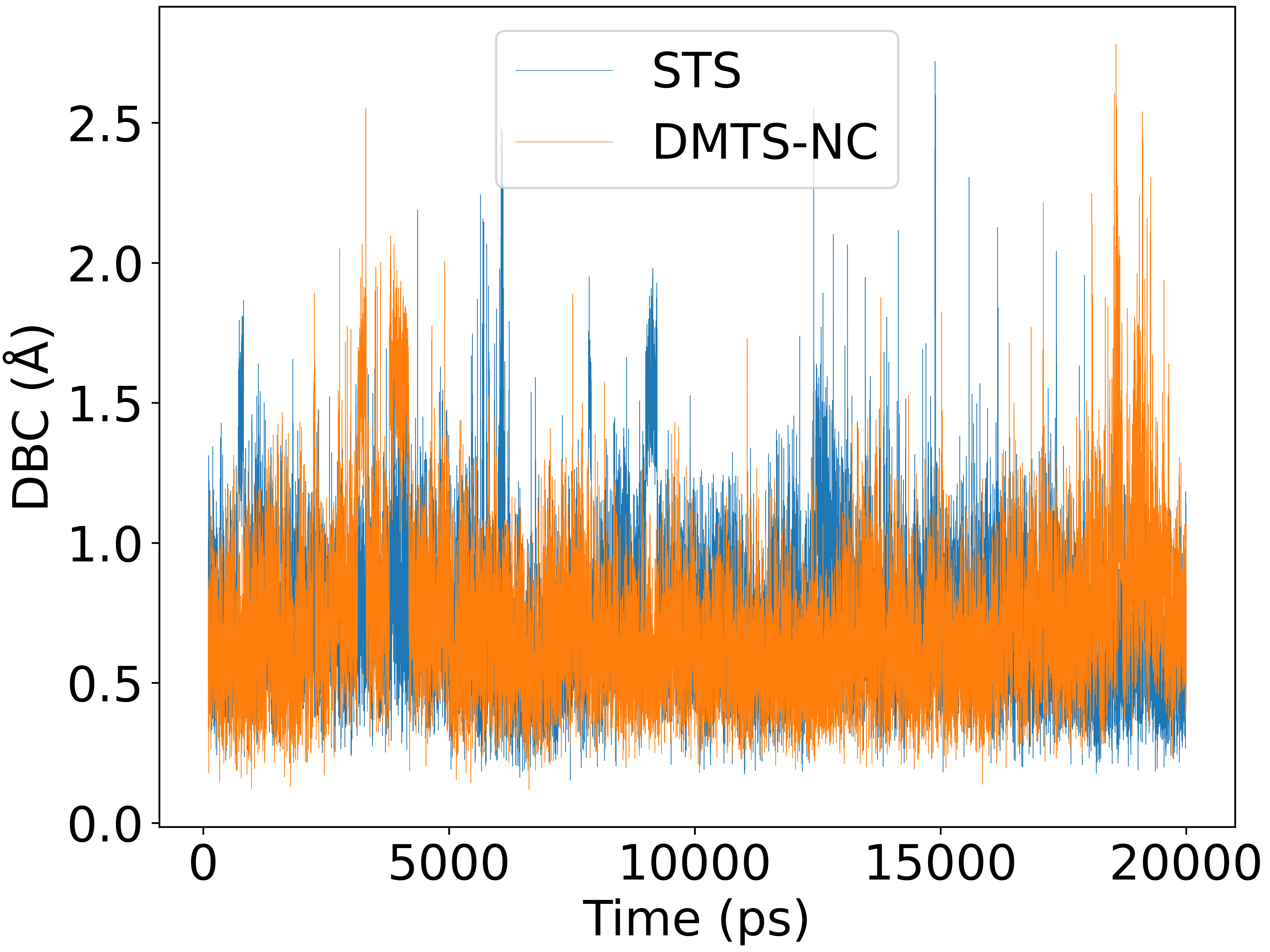}}
    \caption{Time evolution of the protein backbone RMSD (Figure (a)) and the ligand's Distance to Bound Configuraiton (Figure (b)) during a 20ns simulation of the solvated lysosyme-phenol. The DMTS-NC integrator uses $\delta = 1.25$fs--$\Delta = 5$fs time steps.}
    \label{fig:rmsd-dbc}
\end{figure}

Regarding potential energy and average temperature, statistics were gathered from $5$ns of simulation and shown in Table~\ref{table:temp-epot}. Similarly to simulations of bulk water, the low-bias identical-variance deviation of the distributions from STS to DMTS-NC, expected for multi-time-stepping in a stable regime, indicates that using an external time step of $\Delta=5$fs yields stable and accurate simulations, while resonance problems start to arise for larger values of $\Delta$. Finally, in Figure~\ref{fig:rmsd-dbc}, we show the time evolution of the protein backbone RMSD and the ligand's Distance to Bound Configuration (DBC), computed using the Colvars library~\cite{Fiorin2013}, during a 20ns simulation of the solvated phenol-lysosyme complex, with frames collected every picosecond.

\begin{table}[htbp]
\centering
\small

\begin{tabular}{@{} l S[table-format=1.2] S[table-format=1.3] l @{}}
\toprule
& \multicolumn{2}{c}{\textbf{Speed (ns/day)}} & \textbf{Numerical} \\
\cmidrule(lr){2-3}
\textbf{Integrator} & {Phenol-lyso.} & {DHFR} & \textbf{instabilities} \\ 
\midrule
STS                 & 2.55  & 1.85  & No \\
DMTS 1.75-3.5       & 6.59  & 4.46  & No \\
DMTS 1-4            & {NaN} & {NaN} & --- \\
Fine-tuned DMTS 2-4 & 7.45  & 4.97  & No \\
\midrule
DMTS-NC 1-4         & 7.20  & 4.89 & No \\
DMTS-NC 1-5         & 8.42  & 5.57  & No \\
DMTS-NC 1.25-5      & 8.67  & 5.72  & No \\
DMTS-NC 1.1-5.5     & {NaN} & {NaN} & --- \\
\midrule
Max. acc. DMTS-NC/ generic DMTS & {31.6\%} &  {28.3\%} \\
Max. acc. DMTS-NC/ fine-tuned DMTS & {16.4\%} &  {15.1\%} \\
Max. acc. DMTS-NC/ STS & {3.40} &  {3.09} \\

\bottomrule
\end{tabular}
\caption{Speed performances for the two ligand-protein systems. Similarly to bulk water simulations, DMTS-NC allows to use a larger external time step than conservative DMTS, without relying on active learning.}
\label{table:proteinspeed}
\end{table}

\begin{table}[htbp]
\centering
\small
\sisetup{separate-uncertainty, table-align-uncertainty}
\setlength{\tabcolsep}{4pt}
\begin{tabular}{@{} ll l S[table-format=3.2(2)] S[table-format=-1.3(3)] @{}}
\toprule
\multicolumn{2}{@{}l}{\textbf{System}} & \textbf{Integrator} & {\textbf{Temperature}} & {\textbf{Potential energy}} \\
& & & {(K)} & {(kcal/mol/atom)} \\
\midrule
Water boxes & 216 mol. & STS & 299.84 \pm 9.57 & -8.049 \pm 0.042 \\
& & DMTS-NC & 299.87 \pm 9.56 & -8.038 \pm 0.043 \\
\addlinespace
& 1600 mol. & STS & 300.09 \pm 3.60 & -8.047 \pm 0.017 \\
& & DMTS-NC & 299.93 \pm 3.54 & -8.037 \pm 0.016 \\
\addlinespace
& 4000 mol. & STS & 300.61 \pm 2.93 & -8.043 \pm 0.014 \\
& & DMTS-NC & 299.69 \pm 2.22 & -8.039 \pm 0.011 \\
\midrule
Solvated & Phenol-lyso. & STS & 300.15 \pm 1.99 & -7.869 \pm 0.009 \\
proteins & & DMTS-NC & 301.51 \pm 1.99 & -7.850 \pm 0.009 \\
\addlinespace
& DHFR & STS & 299.97 \pm 1.64 & -7.819 \pm 0.008 \\
& & DMTS-NC & 302.26 \pm 1.64 & -7.796 \pm 0.007 \\
\bottomrule
\end{tabular}
\caption{Average temperatures and potential energies ($\pm$ empirical standard deviations). DMTS-NC simulations were performed at the highest stable time steps that we found, i.e. $\delta = 1.25$fs--$\Delta = 5$fs for the protein systems, and $\delta = 1.2$fs--$\Delta = 6$fs for bulk water.}
\label{table:temp-epot}
\end{table}

\paragraph{Hydration free energies of small molecules}

We then compare STS and DMTS-NC for calculating the hydration free energies (HFE) of a collection of 44 solvated small molecules, both in NVT and NPT ensembles. Free energy is computed using the alchemical lambda-ABF method~\cite{Lagardere2024}, following the alchemical parameterization for the FeNNix-Bio1 architecture described in~\cite{Ple2025}.

\begin{figure}[htbp]
    \centering
    \includegraphics[width=\linewidth]{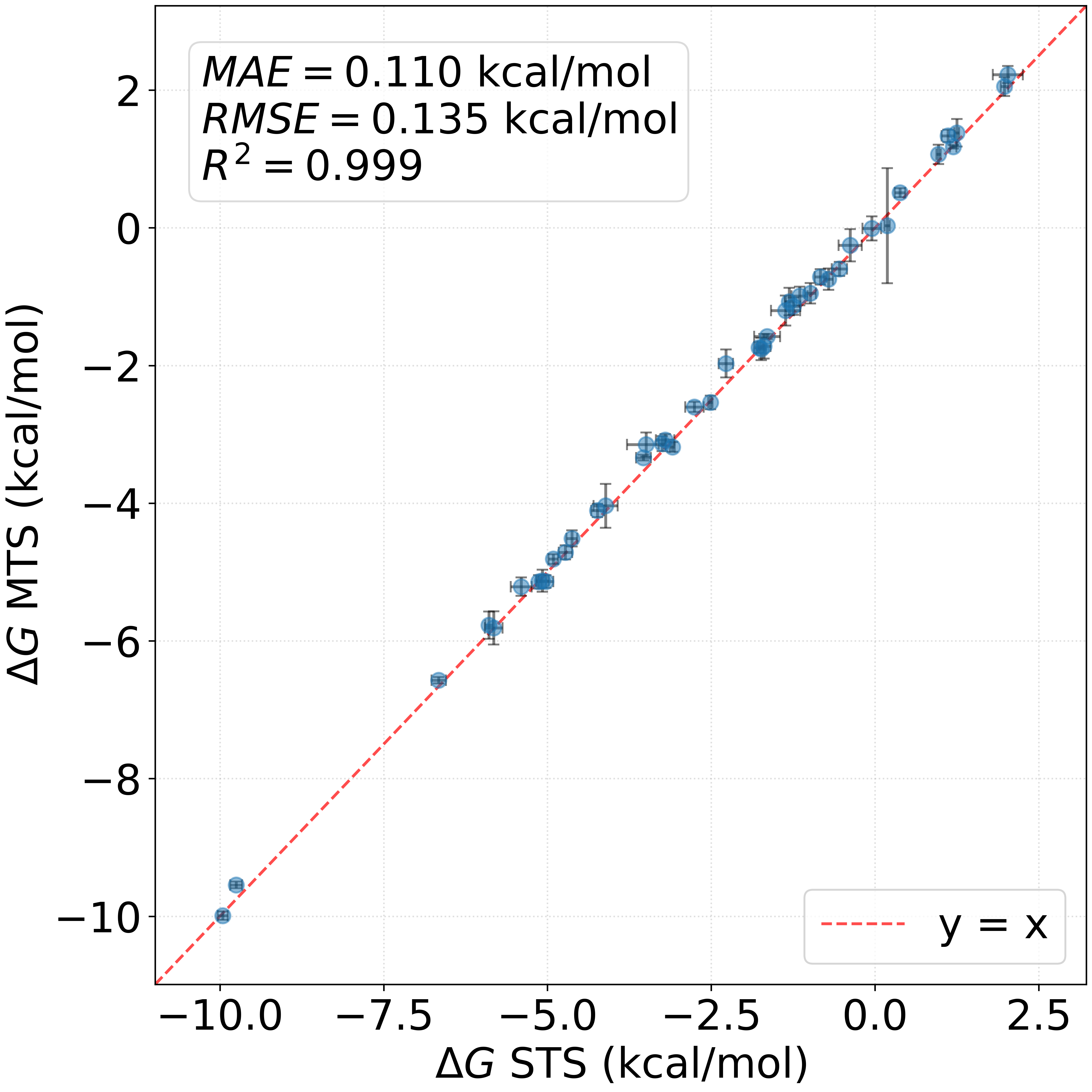}
    \caption{Hydration free energy of 44 small molecules in NPT. Error bars are computed from three independent lambda-ABF runs for each integrator and each molecule.}
    \label{fig:hfe}
\end{figure}

Each simulation was done for $6$ns, with a time step of $1$fs on STS, and $1$fs-$6$fs on DMTS-NC. Results are illustrated in Figure~\ref{fig:hfe}, where error bars result from three runs for each molecule and each setup. Thanks to the increased speed of DMTS-NC, simulation cost is greatly reduced with respect to STS, while preserving its high accuracy: the average mean absolute error between the HFE predicted by STS and DMTS-NC is 0.110 kcal/mol in both NVT and NPT simulations, and the root-mean-square error (RMSE) is 0.142 kcal/mol in NVT and 0.135 kcal/mol in NPT. We refer to Supporting information for a complete list of the molecules, and the corresponding HFE values.

\paragraph{HHF increased speed.}
By damping hydrogen atom movements, the HHF method described in Section~\ref{sec:methods} with a $\gamma_H=10$ps$^{-1}$ friction on hydrogen and $\gamma_O = 1$ps$^{-1}$ on oxygen allowed to push further the external time step to $\Delta = 10$fs on water boxes systems, $\Delta = 7$fs on DHFR, and $\Delta=6$fs on phenol-lysosyme, yielding an additional increase in speed, mostly on bulk water. Simulation speed and accelerations are gathered in Table~\ref{table:speed-hhf}, and statistics on temperature and potential energy are given in Table~\ref{table:temp-epot-hhf}. In bulk water systems, HHF allows for an acceleration between $25.92\%$ and $32.61\%$ with respect to regular DMTS-NC, and from $7.96\%$ to $14.57\%$ on solvated proteins. When comparing with STS, the HHF simulations yield an acceleration factor ranging from $3.45$ (on DHFR) to $5.28$ (on the 4800-atoms water box). While temperature, potential energy, and overall stability are well preserved in this setup, the acceleration provided by HHF comes at the cost of degraded diffusion coefficients, as discussed in Section~\ref{sec:resonances}.

\begin{table}[htbp]
\centering
\small
\begin{tabular}{l l l l l l}
\toprule
& & \multicolumn{2}{l}{Water boxes} & \multicolumn{2}{l}{Solvated prot.} \\
\cmidrule(lr){3-4} \cmidrule(lr){5-6}
\multicolumn{2}{l}{System size / Name} & {1600 mol.} & {4000 mol.} & {Phenol-Lys.} & {DHFR} \\
\midrule
\multirow{3}{*}{Speed (ns/day)} & STS 1 fs       & 9.08  & 3.63  & 2.55  & 1.85 \\
                                & DMTS-NC        & 36.2  & 13.7  & 8.42  & 5.57 \\
                                & HHF            & 47.9  & 17.2  & 9.09  & 6.38 \\
\midrule
\multirow{2}{*}{HHF Accel. w.r.t.}  & DMTS-NC   & 32.61$\%$ & 25.92$\%$ & 7.96$\%$  & 14.57$\%$ \\
                                & STS        & 5.28  & 4.74  & 3.56  & 3.45 \\
\bottomrule
\end{tabular}
\caption{HHF speed performances and acceleration compared to DMTS-NC and STS.}
\label{table:speed-hhf}
\end{table}

\begin{table}[htbp]
\centering
\small
\begin{tabular}{ll c S[table-format=3.0(1)] S[table-format=-1.4(4)] S[table-format=-1.4(4)]}
\toprule
\multicolumn{2}{l}{System} & {Time steps} & {Temperature} & \multicolumn{2}{c}{Pot. Energy (kcal/mol/atom)} \\
\cmidrule(lr){5-6}
& & {(fs)} & {(K)} & {DMTS-NC} & {STS} \\
\midrule
\multirow{2}{*}[-1.2ex]{Water boxes}
& 1600 mol. & 1--10 & 300 \pm 4 & -8.02 \pm 0.02 & -8.05 \pm 0.02 \\
\addlinespace[0.8ex]
\cmidrule(lr){2-6}
\addlinespace[0.8ex]
& 4000 mol. & 1--10 & 300 \pm 2 & -8.02 \pm 0.01 & -8.04 \pm 0.01 \\
\midrule
\multirow{2}{*}[-1.2ex]{Solvated prot.} 
& Phenol-Lys. & 1--6 & 297 \pm 2 & -7.87 \pm 0.01 & -7.87 \pm 0.01 \\
\addlinespace[0.8ex]
\cmidrule(lr){2-6}
\addlinespace[0.8ex]
& DHFR & 1--7 & 298 \pm 2 & -7.81 \pm 0.01
& -7.82 \pm 0.01 \\
\bottomrule
\end{tabular}
\caption{HHF average temperatures and potential energies.}
\label{table:temp-epot-hhf}
\end{table}

\paragraph{Distillation of MACE-OFF23.}

The DMTS and DMTS-NC procedures are generic and can be applied to any NNP. To illustrate this, we applied DMTS-NC to the MACE-OFF23(S) foundation model~\cite{Kovacs2025}. Its distillation (based on the same dataset) into a non-conservative model yielded a good agreement with the data (MAE=$1.484$kcal/mol, RMSE=$2.363$kcal/mol), very similar to the distillation of FeNNix-Bio1(M). We compared speed of simulation between STS and DMTS-NC integrators on the same systems (three water boxes and two solvated proteins), and gathered the results in Table~\ref{table:mace}. Since MACE-OFF is more expensive to evaluate than FeNNix-Bio1 (see~\cite{Ple2025} for a performance comparison), while the distilled non-conservative model has exactly the same architecture as previously, the acceleration factor is greater here than for DMTS based on FeNNix-Bio1, now ranging from $3.66$ to $5.64$.

\begin{table}[htbp]
\centering
\small
\begin{tabular}{l l l l l l l}
\toprule
& & \multicolumn{3}{l}{Water boxes} & \multicolumn{2}{l}{Solvated prot.} \\
\cmidrule(lr){3-5} \cmidrule(lr){6-7}
\multicolumn{2}{l}{System size / Name} & 216 mol. & 1600 mol. & 4000 mol. & Phenol-Lys. & DHFR \\
\midrule
\multirow{2}{*}{Speed (ns/day)} & STS     & 7.40 & 1.04 & 0.38    & 0.31  & 0.22 \\
                                & DMTS-NC & 37.59 & 5.88 & 2.07 & 1.14 & 1.00     \\
\midrule
Acceleration                      &         & 5.08    & 5.64    & 5.45    & 3.66  & 4.62     \\
\bottomrule
\end{tabular}
\caption{DMTS-NC performances with MACE-OFF23(S).}
\label{table:mace}
\end{table}

\subsection{Mitigation of resonances}\label{sec:resonances}

\begin{figure}[htbp]
    \centering
    \includegraphics[width=\linewidth]{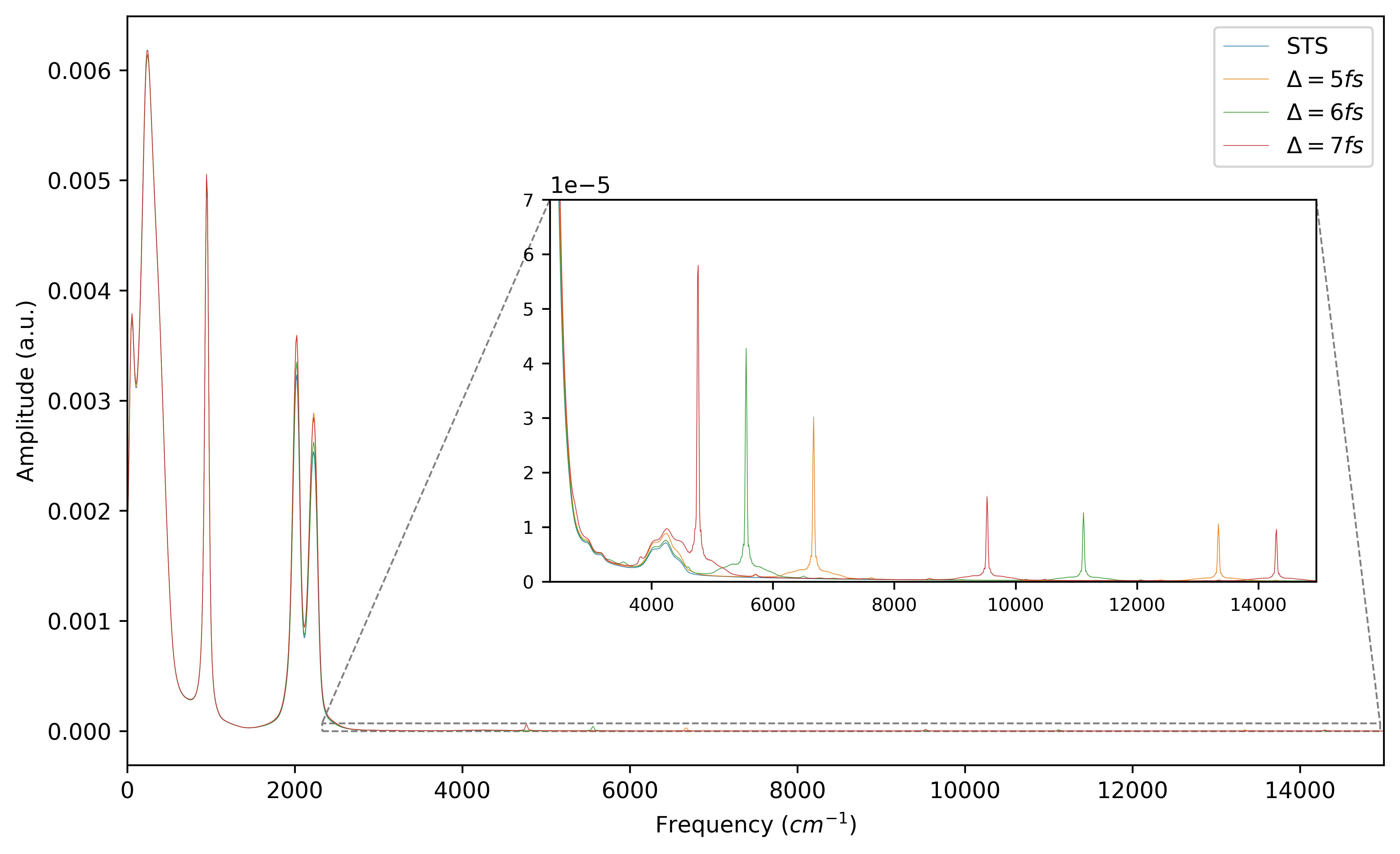}
    \caption{Hydrogen velocity auto-correlation spectra for various external time steps, in a system of 1600 water molecules. From left to right, the first peak corresponds to molecule libration and slow fluctuations in the hydrogen bond network, the second peak to angle bending oscillations and the third and fourth peaks to bond stretching motions. The ones on the inset graph are non-physical artifacts. As the external time step increases, these peaks shift to the left, eventually coupling with the bond-oscillation peak at the stability limit.}
    \label{fig:spectra}
\end{figure}

In MTS methods, the maximum value of the external time step is usually limited by resonance phenomena~\cite{Skeel1993,Skeel2003} resulting from a coupling of internal frequencies of the molecular system with non-physical periodicity created by the various time steps, which can induce numerical instabilities such as abnormal fluctuations of kinetic energy for large values of $\Delta$. This phenomenon is well highlighted on velocity auto-correlation spectra, as shown on Figure~\ref{fig:spectra} (computed from the MD trajectory at the level of the smallest time step using the Wiener-Khinchin theorem) for various external time steps. In this spectral representation, the initial peaks represent physical frequencies, such as intramolecular angle bending (around 1000 cm$^{-1}$) and bond stretching (the two peaks around 2000 cm$^{-1}$). Higher-frequency peaks are non-physical artifacts resulting from MTS. As the external time step $\Delta$ increases, the main artifact shifts towards lower frequencies, eventually coupling with the physical stretching peaks when $\Delta$ becomes too large.

To mitigate this problem and push further the stability limit, various alternative methods have been proposed~\cite{Feenstra1999,Morrone2011,Albaugh2019,Leimkuhler2013c,Margul2016,Gouraud2025}  some of which can be difficult to parametrize, or have a negative impact on certain dynamic properties. As explained in Section~\ref{sec:methods} and illustrated in Section~\ref{sec:perfs}, we explored three such strategies, namely Hydrogen Mass Repartitioning (HMR), High-Hydrogen-Friction (HHF) and Fast Forward Langevin (FFL), all of which are straightforward to implement and to parametrize. When analyzing MTS strategies, or any method aiming at accelerating simulation speed, a standard sanity check consists in computing the diffusion coefficient, a dynamic property that measures the tendency of particles to drift through the system. A significant drop of this quantity can indicate a reduced effective sampling rate, which in turn offsets the computational gain, as longer trajectories become necessary to keep same amount of sampling. As explained in~\cite{Lagardere2019}, the acceleration rate should ideally be higher than the corresponding loss in diffusion. 

\begin{figure}[htbp]
    \centering
    \subfloat[MSD]{\includegraphics[width=0.5\linewidth]{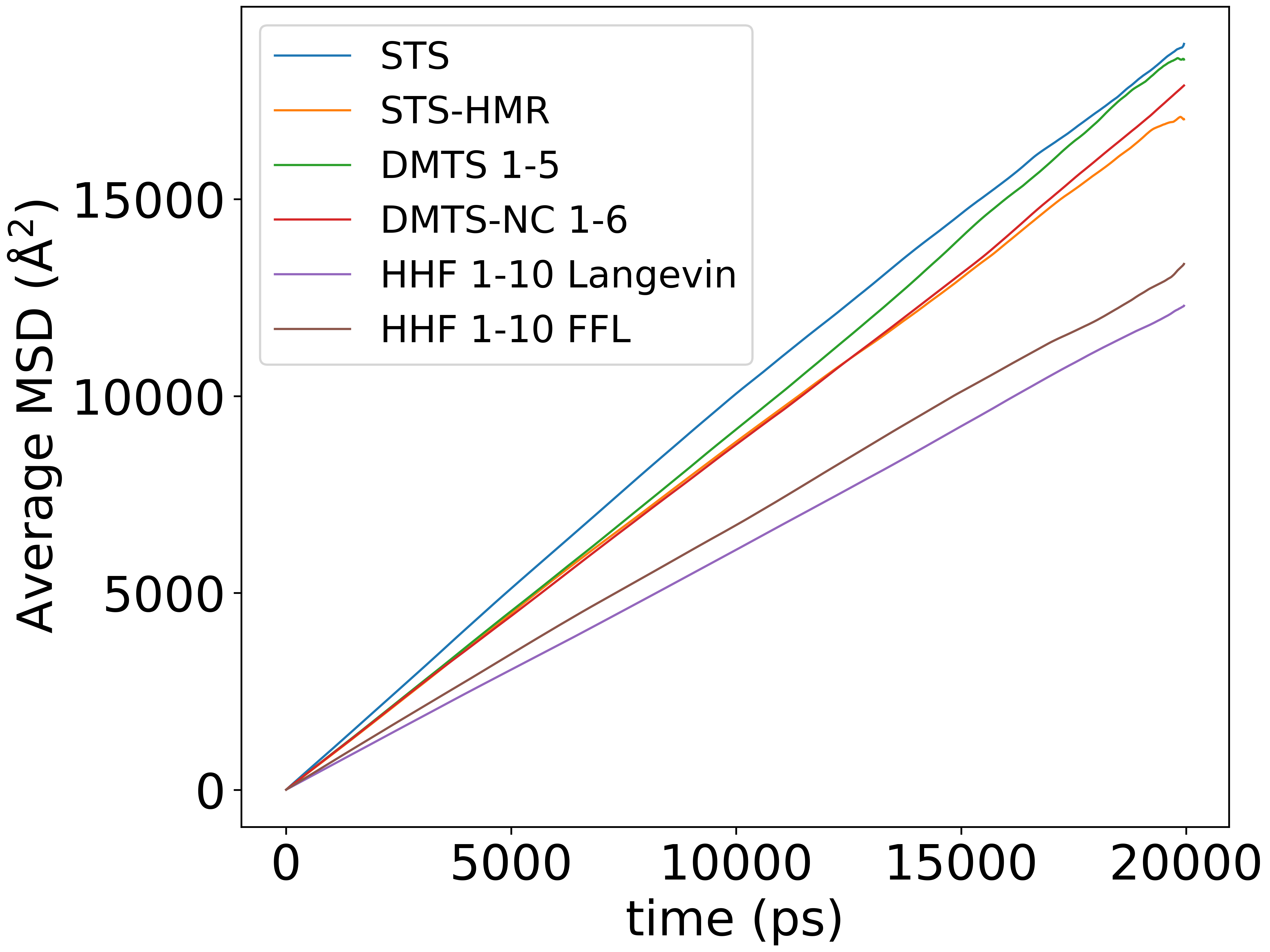}}
    \subfloat[Diffusion coefficient]{\includegraphics[width=0.5\linewidth]{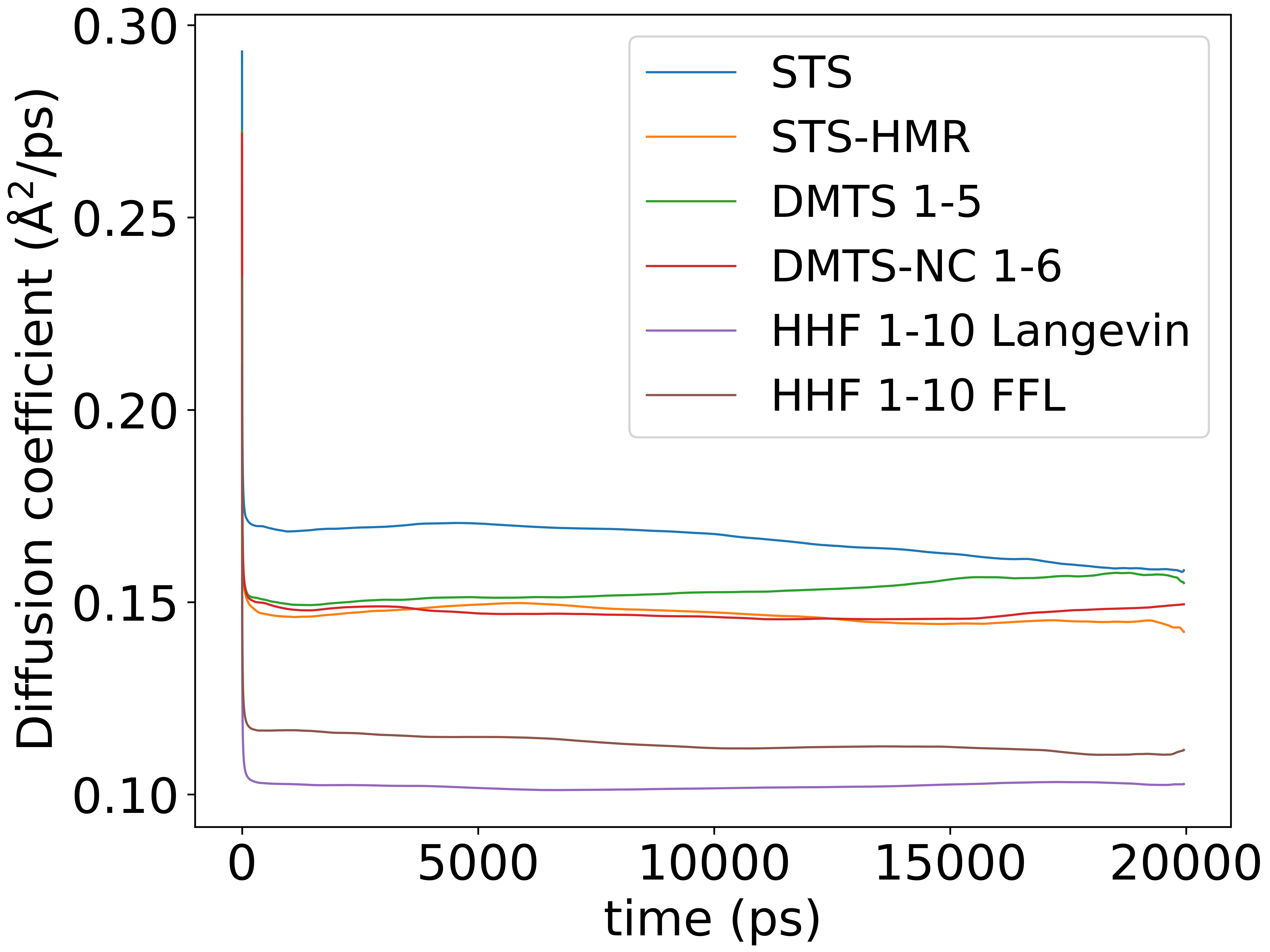}}
    \caption{Mean-square displacement (Figure (a)) and diffusion coefficients (Figure (b)) for various time steps and thermostats.}
    \label{fig:diffusion}
\end{figure}

Here we computed diffusion coefficients using the Einstein formula (see for instance~\cite[Chapter 13]{tuckermanbook}), which relates the diffusion coefficient $D$ to the growth rate of Mean-Square-Displacement (MSD). By denoting $X_t$ the position of the system at time $t$, $N$ the number of atoms and $\E$ the expectation with respect to realizations of the stochastic dynamics, the Einstein formula states that 
\[D = \frac{1}{6N}\lim_{t\rightarrow\infty}\frac{\dd}{\dd t}\E[|X_t-X_0|^2]\,.\]
In other words, the diffusion coefficient is proportional to the long-time limit of the slope of the MSD, which we plot in Figure~\ref{fig:diffusion} for a 4800 atoms water box simulated during 20ns, with a frame saved every picosecond. Quantitative results, gathered in Table~\ref{table:diffusion}, are the following. In all explored setups, the percentage of loss in diffusion with respect to STS is significantly lower than the associated increase in speed: for instance, regular DMTS-NC with $1-6$fs time steps only looses $12.6\%$ in diffusion while accelerating simulation by $4.53$ (see Section~\ref{sec:perfs}). In this case, the reduction in diffusion is largely explained by the use of HMR, as a similar drop is observed ($12.0\%$ loss) by using HMR on the STS integrator. In HHF simulations, as expected, the increased friction on hydrogen atoms resultas in a larger decrease in diffusion, although replacing Langevin thermostat by Fast Forward Langevin thermostat (see Section~\ref{sec:methods}) reduces the loss from $39.2\%$ to $32.6\%$. While the impact on diffusion coefficients is indeed higher in HHF, the acceleration in simulation speed is also greater: on the 4800 atoms water system, HHF simulations with $1-10$fs time steps yield $5.28$-fold acceleration with respect to standard STS. Overall, the combined use of HMR, HHF and FFL helps extending the stability limit of DMTS-NC integrators, with a relatively low impact on diffusion coefficient compared to the associated additional speedup.

\begin{table}[htbp]
\centering
\small
\sisetup{
    scientific-notation = fixed,
    fixed-exponent = -1,
    separate-uncertainty = true,
    table-format = 1.3(2)e-1
}
\begin{tabular}{l S S[table-format=2.1, scientific-notation=false]}
\toprule
Integrator & {Diffusion coeff. (\AA$^2$/ps)} & {Loss w.r.t. STS (\%)} \\
\midrule
STS & 0.1672 \pm 0.0017 & {---} \\
STS-HMR & 0.1472 \pm 0.0012 & 12.0 \\
DMTS 1-5 & 0.1525 \pm 0.0007 & 8.8 \\
DMTS-NC 1-6 & 0.1461 \pm 0.0005 & 12.6 \\
\addlinespace
HHF 1-10 Langevin & 0.1016 \pm 0.0003 & 39.2 \\
HHF 1-10 FFL & 0.1126 \pm 0.0007 & 32.6 \\
\bottomrule
\end{tabular}
\caption{Diffusion coefficients for various MTS strategies. The loss represents the relative decrease in diffusion compared to the STS reference.}
\label{table:diffusion}
\end{table}

\section*{Conclusion} The DMTS-NC approach marks an improvement in atomistic molecular dynamics with neural network potentials by seamlessly merging model distillation with non-conservative (NC) forces and multi-time-stepping (MTS) strategies. By leveraging a distilled architecture that strictly enforces physical priors—such as equivariance under rotation and the cancellation of atomic force components—this method secures a 15-30\% speedup over its conservative predecessors and provides up to a 5.6-fold acceleration over single-time-step methods. Remarkably, the framework remains architecture-agnostic and robust, delivering near-ab initio accuracy across foundation models like FeNNix-Biol and MACE-OFF23 without requiring the labor-intensive burden of system-specific fine-tuning. When integrated with stability-extending techniques like Hydrogen Mass Repartitioning and High Hydrogen Friction, DMTS-NC successfully narrows the long-standing performance gap between high-accuracy neural network potentials and efficient classical force fields, setting a new standard for high-throughput, reliable simulations.These advancements represent a further step toward closing the performance gap with classical force fields. Future work will focus on optimizing our implementation to handle more complex systems and investigate subtle sampling properties. As a future direction, we plan to explore further acceleration frameworks, including splitting schemes inspired by RESPA-1 and RESPA-2~\cite{Zhou2001,Lagardere2019} as well as various resonance-mitigating strategies such as isokinetic phase-space algorithm~\cite{Leimkuhler2013c} and velocity jumps~\cite{Gouraud2025}.

\section*{Competing interests}

Louis Lagardère and Jean-Philip Piquemal are shareholders and co-founders of Qubit
Pharmaceuticals. All the remaining authors declare no conflict of interest.

\section*{Authors Contribution Statement}
NG, TP, LL, JPP conceived and designed the research and numerical experiments. NG performed the numerical experiments using FeNNix-Bio1. C. C performed the distillation and tests using MACE-OFF23. NG and OA wrote the code. All authors analyzed the data. NG, TP, and JPP wrote the paper with the input of all other authors.

\section*{Acknowledgement}

This work has received funding from the European Research Council (ERC) under the European Union's Horizon 2020 research and innovation program (grant agreement No 810367), project EMC2 (JPP). Computations have been performed at IDRIS (Jean Zay) on GENCI Grants: n° A0150712052 (J.-P. P.) and A0180716167 (N.G.).

\section*{Code availability}

Calculations were performed using the FeNNol library:~\url{https://github.com/FeNNol-tools/FeNNol} and the Tinker-HP package~\url{https://github.com/TinkerTools/tinker-hp}. Distance to Bound Configuration (DBC) trajectories were computed using the Colvars library~\url{https://github.com/Colvars/colvars}.

\section*{Data availability}

Pretrained models are available on Github at:~\url{https://github.com/FeNNol-tools/FeNNol-PMC}.
\section*{Supporting Information}
Pressure and density distributions;
Free energies of solvation: list of studied molecules (NVT and NPT).

\printbibliography

\newpage
\section*{TOC Graphic}
\begin{figure}
    \centering
    \includegraphics[width=0.9\linewidth]{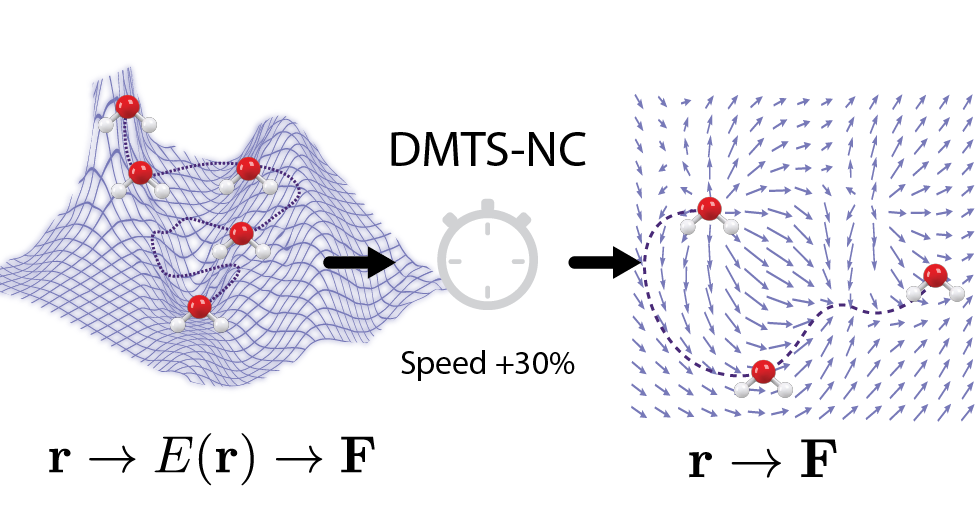}
    \caption*{For Table of Contents Only}
\end{figure}

\end{document}